\documentclass[12pt,preprint]{aastex}
\usepackage{amsmath}
\usepackage{amssymb}
\usepackage{graphicx}

\usepackage{natbib}
\def\aap{A\&A} 
\def\apj{ApJ} 
\def\apjl{ApJL} 
\def\araa{ARAA} 
\def\mnras{MNRAS} 
\def\nat{Nature} 

\title{The Different Structures of the Two Classes of Starless Cores}

\author{Eric Keto\altaffilmark{1} and
Paola Caselli \altaffilmark{2}}
\affil{Harvard-Smithsonian Center for Astrophysics, 60 Garden
  Street, Cambridge, MA 02138, USA}
\affil{School of Physics and Astronomy, University of Leeds, Leeds LS2 9JT}
\altaffiltext{1}{keto@cfa.harvard.edu}
\altaffiltext{2}{p.caselli@leeds.ac.uk}

\shorttitle{Starless Cores}
\shortauthors{Keto and Caselli}

\begin{document}

\begin{abstract}

We describe a model for the thermal
and dynamical equilibrium of starless cores that 
includes the radiative transfer of the gas and
dust and simple CO chemistry.
The model shows that
the structure and behavior of the cores is significantly different
depending on whether the central density is either above
or below about $10^5$ cm$^{-3}$. This density is significant as the
critical density for gas cooling by gas-dust collisions
and also as the critical density for dynamical stability, 
given the typical properties of the
starless cores. 
The starless cores thus divide into two classes that
we refer to as thermally super-critical and thermally sub-critical.
This two-class distinction allows an improved interpretation of 
the different observational data of starless cores within a single model.
\end{abstract}


\section{Introduction}

Starless cores
are dense regions ($n_{\rm H_2} \sim 10^4$ to $10^6$ cm$^{-3}$) in dark clouds
with linear scales of tenths of pc, and total masses of a few solar masses. The
starless cores contain no infrared sources above the sensitivity level of
the IRAS satellite (about 0.1 L$_\odot$ at the distance of Taurus) and thus are
thought to be sites of possible future rather than current star formation
\citep{MyersLinkeBenson1983, MyersBenson1983, BensonMyers1989, Beichman1986, Ward-Thompson1994,
Tafalla1998, LeeMyers1999, Ward-ThompsonMotteAndre1999, Bacmann2000, Shirley2000, 
LeeMyersTafalla2001, DiFrancesco2007, BerginTafalla2007}. 
Detailed observations of individual starless cores \citep{AlvesLadaLada2001,Tafalla2004} 
show density
profiles that approximately match those
of pressure-confined, hydrostatic spheres
\citep[Bonnor-Ebert or BE spheres; ][]{Bonnor1956}.

Despite an overall similarity in structure,
recent observations 
with increasingly better angular resolution
and sensitivity show that the starless cores are
not all the same. For example, observations of L1517B
indicate
isothermal gas and show spectral line profiles indicative of little motion or
possibly expansion 
\citep{Tafalla2004, Keto2004, Sohn2007}. 
In contrast, observations of L1544 indicate a significant variation in temperature from
center to edge \citep{Crapsi2007} and show spectral line profiles
consistent with an overall contraction  \citep{Williams1999, Caselli2002, Keto2004, Sohn2007}. 
The cores B68 and TMC-1 provide
another example in contrast. Observations of CO in B68 indicate
an excitation temperature rising  toward the edge of the core 
\citep{Bergin2006},
and show spectral line profiles consistent with
internal oscillations (sound waves) \citep{Lada2003}.
In comparison, observations of CO isotopologues in TMC-1C 
 \citep{Schnee2007} indicate an
excitation temperature decreasing  toward the 
edge and show spectral line profiles consistent with contraction.

In an earlier paper \citep{KetoField2005} we suggested that  the
observed
differences were consistent with the different
temperature and density 
structures of starless cores that have
central densities greater or less than about $10^5$ cm$^{-3}$.
The significance of this particular density is twofold. 

First, this
density divides the starless cores into those that are dynamically
stable (those with lower densities) and those that must be contracting
or collapsing (those with higher densities). The 
dynamical stability, of course, also depends on the total mass and 
the internal energy of the core, but
for starless cores with masses of 1 - 10 M$_\odot$, 
temperatures of around 10 K, and subsonic internal velocities,
the critical density for dynamical 
stability falls within a range around $10^5$ cm$^{-3}$.  

Second, at densities below a few 10$^5$ cm$^{-3}$
the molecular
gas in starless cores cools primarily through molecular line
radiation, but  at higher densities the gas cools
primarily through collisional coupling with the dust.   
This difference in cooling results in different temperature structures
in cores with central densities below or above  $10^5$ cm$^{-3}$.
If the density in a
core is everywhere below $10^5$ cm$^{-3}$ so that the gas and 
dust are nowhere 
collisionally coupled, then
the entire core cools primarily by molecular line radiation. 
As a result, the core is approximately isothermal.
If the core has a central density above $10^5$ cm$^{-3}$, the
gas in the core center cools by collisional coupling with the
dust  whereas the gas elsewhere cools by 
molecular line radiation. This difference
in cooling results in a variation in
temperature across the core. Because the dust, heated by
external starlight, is cooler in the core center, the gas there,
coupled to the dust,  is also cooler.

In this paper we continue our investigation of the
structures of the two classes of starless cores. We
improve our previous model of thermal equilibrium by accounting for the
heating of the gas by hot electrons photoelectrically released
from dust grains and by including variable molecular abundances
based on a simple model for CO chemistry. Our improved model for 
starless cores allows a
better interpretation of the observational data. 

For example,
the inclusion of photoelectric heating provides a
better description of the thermal structure near the core boundary
where this effect  is  most significant.
Applied  
to recent observations of the edges 
of the starless cores B68 and TMC-1C \citep{Bergin2006, Schnee2007}
the rising and falling CO excitation temperature seen in the observations 
of the two cores, respectively, are consistent with the different effect 
of photoelectric heating on cores with relatively higher and lower
densities at their boundaries. 

The loss from the gas phase 
of CO and other molecular coolants has only a minor 
effect on the gas temperature and hence the density structure of starless
cores \citep{Goldsmith2001}. Nonetheless, an understanding of the molecular 
abundances is quite important in the interpretation of spectral line
observations, particularly the line profiles and strengths. We have used
our simple model of CO chemistry in modelling spectral line
profiles in an oscillating lower mass, lower density
starless core \citep{Broderick2007} and in
modelling spectral line strengths in a collapsing, higher mass,
higher density, starless core
(E.~Keto \& P.~Caselli, in prep.).

Our improved models of starless cores are also useful in constraining some
of the properties of the interstellar medium (ISM) that relate to the
thermal equilibrium of the starless cores.
In particular,  we obtain temperatures closer to those indicated by
observations with dust opacities higher
than those of naked grains and
more consistent 
with fluffy dust \citep{OssenkopfHenning1994, KruegelSiebenmorgen1994}, and
cosmic-ray ionization rate towards the lower end of the range of
estimates, $1\times 10^{-17}$ s$^{-1}$  to $6\times 10^{-17}$ s$^{-1}$.  
We find that photoelectric heating  
improves the stability of cores by increasing the confining pressure with reduced
overlying mass, but additional internal energy is still required
if the larger cores are to be supported.

\section{The two classes of starless cores} \label{twoCategories}

\subsection{Observational description}

The higher mass
and higher density cores are unstable to
collapse and the formation of protostars within a free-fall time,
but being starless, these cores have not yet formed a protostar.
Thus these cores may be thought of as "young brides" that are soon to give
birth to stars. The density profiles within these cores are more centrally 
concentrated owing to the greater importance of gravitational forces
in confining the core.
These cores often show asymmetric spectral line profiles \citep{Williams1999, 
Gregersen1997, Launhardt1998, LeeMyersTafalla1999,
GregersenEvans2000, Caselli2002, Keto2004,
LeeMyersPlume2004a, LeeMyersPlume2004b, Sohn2004} consistent with
inward motions and line widths that increase toward the centers of
the cores consistent with inward acceleration of the infall. 
Spectral lines from some cores show two components separated by more
than the sound speed indicating supersonic velocities
\citep{Sohn2007}.
These denser cores are not isothermal. Observations of both dust 
\citep{WTAndreKirk2002, Pagani2003, 
Pagani2004, Schnee2005} and gas \citep{Crapsi2007,  Pagani2007} 
indicate cooler temperatures
in their centers, the latter consistent with the change from molecular
line cooling to more efficient dust cooling as the 
density increases toward the interior.

In contrast, lower mass and lower density cores are stable against gravitational
contraction, and therefore
in the absence of changes in their environment, will remain
starless.  These barren cores may be thought of as "old maids".  
These cores have 
flatter density profiles consistent with the relatively greater importance
of an external pressure in confining the cores. Their spectral
line profiles may be simple Gaussians indicating little internal
motion or show complex 
shapes consistent with internal oscillations (sound waves) 
\citep{Lada2003, Redman2006, Aguti2007, Maret2007}.  
Multi-wavelength infrared observations also indicate internal
density perturbations
\citep{Steinacker2005a, Steinacker2005b}. 
The linewidths are generally more constant across the core and indicate
subsonic internal velocities. These cores are more nearly isothermal ($\sim 10$ K)
consistent with molecular line cooling throughout.

Some recent papers have also drawn attention to differences within the 
population of starless cores and use terms such as "pre-stellar" to indicate a 
tendency
toward collapse and star formation versus "starless" or "failed" to suggest
the opposite \citep{Andre2008}. Sometimes the distinction is drawn as
"gravitationally bound" and "gravitationally unbound"
suggesting a
ratio of greater or less than unity of 
gravitational energy to internal energy including thermal energy
and possibly microturbulence and magnetic energy. These  
terms suggest a division of the starless cores into two classes distinguished
by their gravitational stability and future evolution. \citet{KirkWTAndre2005}
divide the starless cores into two categories called "bright" and "intermediate".
These refer to the brightness of the dust emission which is related to
the mass and density of the cores.
In this paper we suggest that 
the distinction is motivated by other observational properties as well, and that
the various properties of the two classes may be derived from a
single theoretical model for the starless cores that we describe in the following
several sections.

\subsection{Theoretical description: Thermally Super-Critical and Thermally Sub-Critical}
We refer to the two classes of cores as ``thermally super-critical'' and ``thermally
sub-critical''. We choose these terms to emphasize that the distinction is
based on the physics of the cores rather then their origin or evolution. The 
term ``thermal''
in this description is appropriate because it refers to both their thermal-radiative 
equilibrium and also to their thermally-supported dynamical equilibrium.
Whether the cores are supported primarily by thermal energy or by magnetic
energy is still a matter of debate. The evidence is 
gathering on the side of thermal support with non-thermal
energy being an important but minor contributor to their stability
\citep{BarrancoGoodman1998, Goodman1998, Tafalla2004, Lada2008}.
In other words, without any non-thermal energy most of the cores would be
unstable. However, to achieve stability, 
the cores require only an amount of  non-thermal 
energy that is a fraction of their thermal energy. The terms thermally super and
sub-critical are similar in meaning to the terms
magnetically super and sub-critical that are well established in the literature,
but the "thermal" modifier suggests that thermal energy is here more important
than magnetic energy. It is
not necessary to prove this supposition to adopt the terminology, and in any
case, in this study we model the cores as thermally supported.

As typical parameters for our calculations, we adopt
a central density of $10^6$ cm$^{-3}$ and
a total mass of 10 M$_\odot$ to represent a thermally super-critical core,
and $10^5$ cm$^{-3}$ and
M$_\odot$ for a thermally sub-critical core. 
The well-studied cores L1544 and B68 are estimated to have
similar central densities.
 \citet{Crapsi2007}
estimate $2\times 10^6$ cm$^{-3}$
for L1544 and \citet{Keto2006} and \citet{Bergin2006} estimate
 $2.5\times 10^5$ cm$^{-3}$ and $3.0 \times 10^5$ cm$^{-3}$ 
respectively for B68, all three estimates
based on fitting isothermal BE spheres to the observational
data. While L1544 and B68 are prototypes of cores in the two classes, 
here we are not attempting to model these two cores exactly. 
Our example parameters are chosen as round numbers that
provide a factor of ten difference in density and mass, 
sufficient to bring out the differences in their 
structures. Of course, the real cores have combinations of mass and
density that form a continuum that includes these two examples.
Other parameters for the cores are listed in table 1.

\section{CO chemistry}

Molecular abundances in starless cores are the result of complex interactions involving
gas-phase chemistry, freeze-out and desorption, and photodissociation. In order to
develop a simplified model we assume that 
the abundance
of CO is decreased by two effects:
1) freeze-out onto dust grains in the high density center, and 2) 
photodissociation at the edge of the core.

\subsection{The photodissociation region}

\begin{figure}[t]
\includegraphics[width=6in]{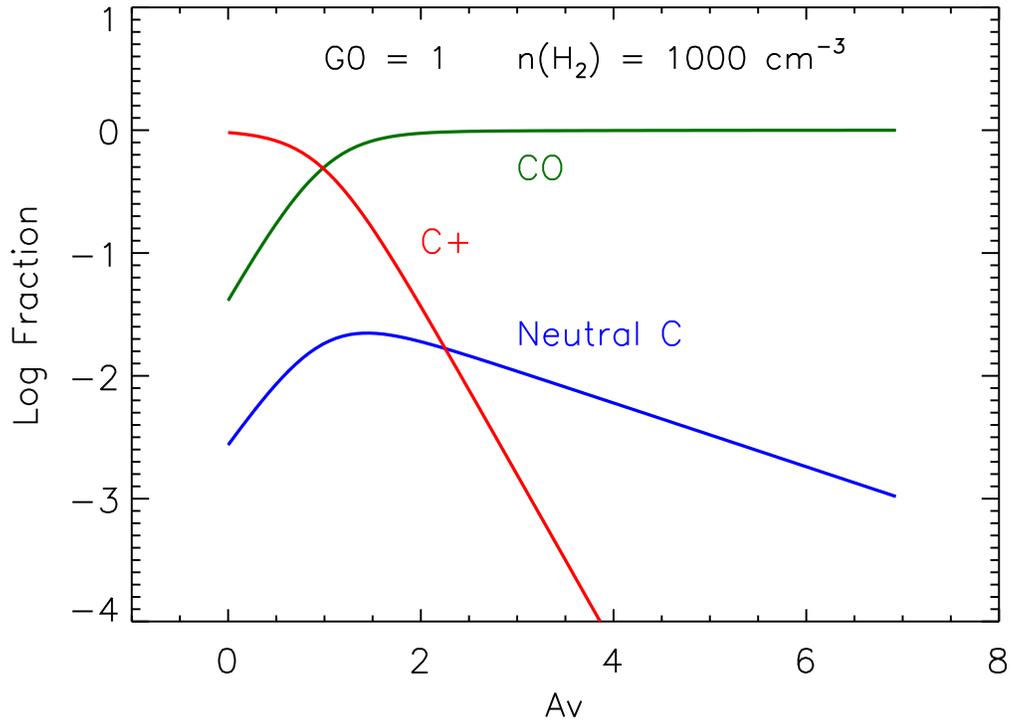} 
\caption{ 
Variation in the abundances of C, CO, and C$^+$ in the photodissociation region
at the edge of a cloud as calculated by the 
simple model described in the text. In this figure, the cloud has a constant 
density of 1000 cm$^{-3}$ and the interstellar radiation field has a Habing
flux, $G_0 = 1$.
} 
\label{fig:co_abundance}
\end{figure}

\citet{TielensHollenbach1985} suggest that at the edge of a molecular
cloud where photodissociation is
important, the dominant cycle for the formation and destruction of CO is,
\def\mapright#1{\quad\smash{\mathop{\longrightarrow}\limits^{#1}}\quad}
\begin{equation}
{\rm C}^+ { \mapright{\rm H_2} } {\rm CH}_2 {\mapright{\rm O} } 
{\rm CO} { \mapright{h\nu} } {\rm C} {\mapright{h\nu} } {\rm C}^+
\end{equation}
The time scale for the formation of CH$_2$ by radiative association is longer than for the
formation of CO from CH$_2$ and O. Therefore we assume that the rate of formation of CO is
given by that of CH$_2$. This eliminates consideration of the complex chemistry of oxygen.
The CO cycle may then be described by three rate equations
for the creation and destruction of C$^+$, C, and CO, for example,
\begin{equation}
{{d{\rm C}^+}\over{dt}} = -R_D({\rm C^+} )  +R_C({\rm C^+} )
\end{equation}
where $R_D$ and $R_C$ are the rates for the destruction and creation of C$^+$.
Similar equations hold for C and CO. The conservation equation in the form, 
\begin{equation}
{\rm C}^+ + {\rm CO} + {\rm C} = 1
\end{equation}
indicates that the symbols stand for the non-dimensional relative abundance 
of the three species
with respect to the total abundance of carbon.
Assuming steady state we may solve for the relative abundance of CO as,
\begin{equation}
{\rm CO} = \bigg( {{\rm CO}\over{\rm C^+}} \bigg) 
	\bigg( 1 + {{\rm CO}\over{\rm C^+}} + {{\rm C}\over{\rm C^+}} \bigg)^{-1}
\end{equation}
The ratios of the relative abundances, ${\rm CO}/{\rm C^+} $ 
and 
$ {\rm C}/{\rm C^+}$, are equal to the ratios of the
rates of destruction and creation of the respective species in the numerators. 
Using the data in 
\citet{TielensHollenbach1985}, these ratios are,
\begin{equation}
{{\rm CO}\over{\rm C^+}} = 
	{{ 1.4\times 10^{-11} G_0 \exp({-3.2 A_V}) }\over
	{ 2.1\times 10^{-10} G_0 \exp({-2.6 A_V}) }}
\end{equation}
and
\begin{equation}
{{\rm C}\over{\rm C^+}} = 
	{{ n({\rm H_2}) 6\times 10^{-16} } \over
	{ 2.1\times 10^{-10} G_0 \exp({-2.6 A_V}) }}
\end{equation}
Here, $n({\rm H_2})$ is the number density of hydrogen, $G_0$ is the interstellar
radiation field in units of the Habing flux (Habing 1968), $A_V$ is the mean
visual extinction to the interstellar radiation field, and the numerical values of the
coefficients are
taken from tables 5 and 12 of \citet{TielensHollenbach1985}.  
The mean extinction at each point in the cloud is computed as the 
extinction from the point to the cloud surface averaged over all
directions,
\begin{equation}
\langle \exp(-A_V) \rangle = {{1}\over{4\pi}}\int \exp(-A_V) d\Omega
\end{equation}
The Habing flux
$G_0 = 1$ corresponds to the average interstellar radiation field of the Galaxy.
Figure \ref{fig:co_abundance}
shows the variation of the three species, CO, C, and C$^+$ as a
function of $A_V$ for a cloud with a constant density of 1000 cm$^{-3}$ and 
$G_0 = 1$. For this flux level, the simple model provides for exponential replacement
of CO and C$^+$ at an $A_V$ of about 1. The more complex photodissociation
model of \citet{TielensHollenbach1985} shows that this exponential replacement
is the dominant relationship,
although in their figure 9b the replacement occurs at a higher $A_V$ of $3-4$ 
because of the higher radiative flux in their model,  $G_0=10^5$.
 
\subsection{Freeze-out}

By comparing dust and molecular line emission, 
\citet{WillacyLangerVelusamy1998},  
\citet{Caselli1999}, 
\citet{Bergin2001},
\citet{Bacmann2002},
\citet{Bacmann2003},
\citet{Caselli2002}, 
\citet{Hotzel2002},
\citet{Bergin2002}, 
\citet{Tafalla2002}, 
\citet{Redman2002}, 
\citet{Tafalla2004}, 
\citet{Crapsi2004}, 
\citet{Crapsi2005}, 
\citet{Pagani2005},
\citet{Tafalla2006}, 
\citet{Schnee2007}, and
\citet{Carolan2008}
were able to measure variations
in the gas-phase abundance of several molecules in several cold, dark clouds.
The observations suggest that the molecular abundance in dense gas is an equilibrium
between the rate of depletion from the
gas phase as molecules freeze onto dust grains and the rate of the inverse process of
desorption
\citep{
BrownCharnleyMillar1988,  
HasegawaHerbstLeung1992,  
BerginLangerGoldsmith1995, 
Aikawa2001, 
Li2002, 
AikawaOhashiHerbst2003, 
Pavlyuchenkov2003, 
Shematovich2003, 
LeeBerginEvans2004, 
Aikawa2005}. 
This suggests that the steady state abundance of 
CO in dense gas may be computed as
the ratio of the depletion time to the sum of the depletion and desorption
times. 

The time scale for depletion onto dust may be estimated as \citep{Rawlins1992},
\begin{equation}
\tau_{on} = (S_0 R_{dg} n({\rm H_2}) \sigma V_T)^{-1}
\end{equation}
Here $S_0$ is the sticking coefficient, with $S_0=1$ meaning that the molecule sticks
to the dust in each collision; $R_{dg}$ is the ratio of the number density of dust grains 
relative to molecular hydrogen; $\sigma$ is the mean cross-section of the dust grains;
and $V_T$ is the relative velocity between the grains and the gas. 
If the grains have  a power law distribution of sizes with
the number of grains of each size scaling as the -3.5 power of their 
radii \citep{MathisRumplNordsieck1977}, then we can estimate their mean cross-section
as,
\begin{equation}
\langle \sigma \rangle = \bigg(\int^{a_2}_{a_1} n(a)da\bigg)^{-1}\int^{a_2}_{a_1} n(a) \sigma (a) da
\end{equation}
where $a_1$ and $a_2$ are the minimum and maximum grain sizes. If $a_1 = 0.005$ $\mu$m 
and $a_2 = 0.3$ $\mu$m, then $\langle \sigma \rangle = 3.4\times 10^{-4}$ $\mu{\rm m}^2$.
Similarly, the ratio of the number densities of dust and gas may be estimated by
computing the mean mass of a dust grain and assuming the standard gas to dust mass
ratio of 100. If the density of the dust is 2 grams cm$^{-3}$, then the ratio of
number densities is $R_{dg} = 4\times 10^{-10}$. 
The relative velocity due to thermal motion is,
\begin{equation}
V_T = \bigg ( {{8kT}\over{\pi\mu}}      \bigg)^{-1/2}
\end{equation}
where $T$ is the temperature and $\mu$ the molecular weight.

There are several processes that generate heat on dust grains
that can liberate frozen molecules to the gas phase.
These include exothermic
H$_2$ formation on the grains, cosmic ray collisions with dust
grains, and photo-desorption of molecules off the grains \citep{WillacyWilliams1993}.
The UV radiation for this latter process derives from the ionization of the
molecular gas by cosmic rays.
According to \citet{Roberts2007}, in starless cores 
these processes are all independent of  density. The rate 
of H$_2$ formation is independent of the molecular density because 
the density of HI  in starless cores is independent of the molecular density. 
The rates of all these desorption processes are not well defined,
nor is it known which is dominant.  Therefore, for
definiteness we
adopt the rate given by \citet{HasegawaHerbst1993} for desorption by cosmic rays
but with the understanding that other processes might also be important.
The time scale is,
\begin{equation}
\tau_{off} = 3.3\times 10^6 \bigg( {{ \zeta }\over{10^{-17}}} \bigg)^{-1} \exp(E_{\rm CO}/70)
{\rm \ \ (yrs)}
\end{equation}
where the cosmic-ray ionization rate $\zeta = 3\times 10^{-17}$ s$^{-1}$, 
and the binding energy
of CO onto ice is $E_{\rm CO} = 1100$ K \citep{Oberg2005, Oberg2007}. 
We choose this value for the binding energy because it is between
the binding energies of CO on water ice, 1180~K 
\citep{Collins2003a,Collins2003b}, and on CO ice,
850~K \citep{Fraser2001}, assuming a typical solid H$_2$O/CO abundance 
ratio of $\simeq$30 on ice 
mantles in dense clouds \citep{Whittet2007}. 

\begin{figure}[t]
\includegraphics[width=6in]{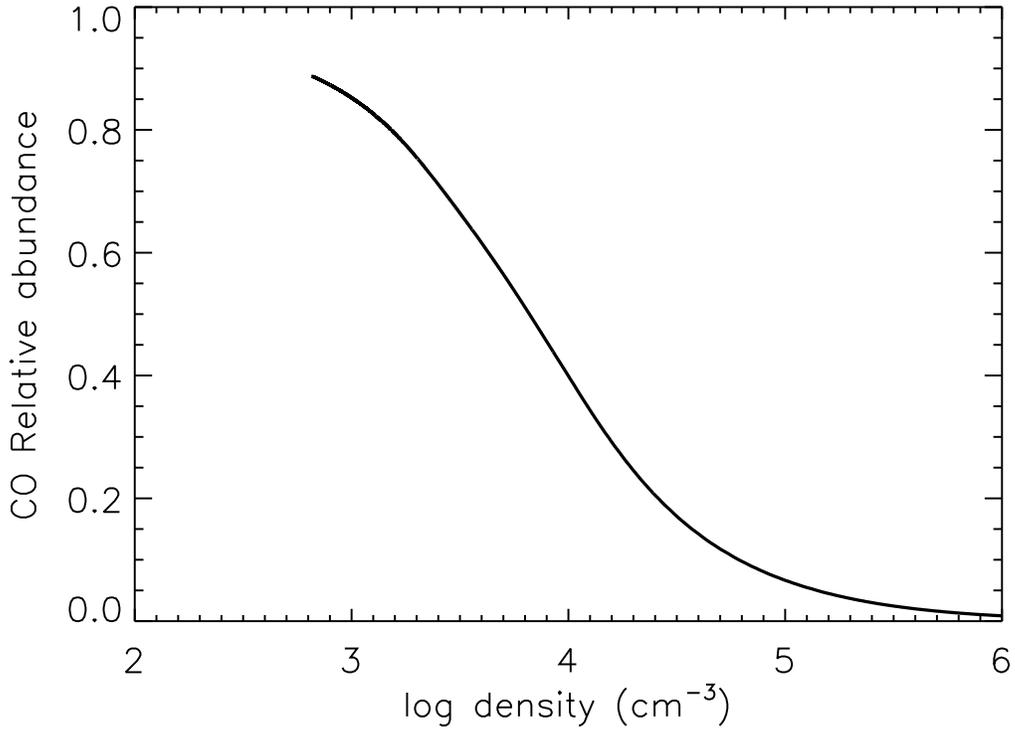} 
\caption{ 
The relative abundance of CO in the gas phase in equilibrium between depletion
and desorption as a function of density. The depletion is caused by the loss of 
CO from the gas phase as the molecules freeze onto dust grains
(equation \ref{eqn:depletion} in text). The depletion is a function of
density because the freeze-out is proportional to the dust-gas collision rate and
inversely proportional to the cosmic-ray induced desorption rate which is itself
independent of density.
} 
\label{fig:depletion}
\end{figure}

The relative abundance of CO in the gas phase 
in equilbrium between depletion and desorption  is,
\begin{equation}\label{eqn:depletion}
	{\rm CO}_{gas} = {{\tau_{on}}\over{\tau_{on} + \tau_{off}}}
\end{equation}
where ${\rm CO}_{gas}+ {\rm CO}_{grains} = 1$, and the symbols stand
for the non-dimensional relative abundance of CO in the two phases
with respect to the total CO abundance.
The depletion time scales inversely with the collision rate while the desorption time
scales inversely with the cosmic-ray ionization rate. Therefore
the depletion of CO is dependent only on
the gas density because the cosmic-ray ionization rate is assumed to 
be the same throughout the cloud.
Figure \ref{fig:depletion} shows the dependence of
the steady state abundance of CO versus the molecular gas density.

While we only illustrate the steady state abundance of CO in this paper, the model
for depletion and desorption may also be used to describe the time dependent 
evolution of CO between the gas and grain phases. The change with time of the abundance in
the gas phase is,
\begin{equation}
{{d \rm CO}_{gas}\over{dt}} = -{{ {\rm CO}_{gas}}\over{\tau_{on}}} 
	+ {{ {\rm CO}_{grain}}\over{\tau_{off}}}
\end{equation}

\subsection{CO abundances in model cores}

\begin{figure}[t]
\includegraphics[width=6in]{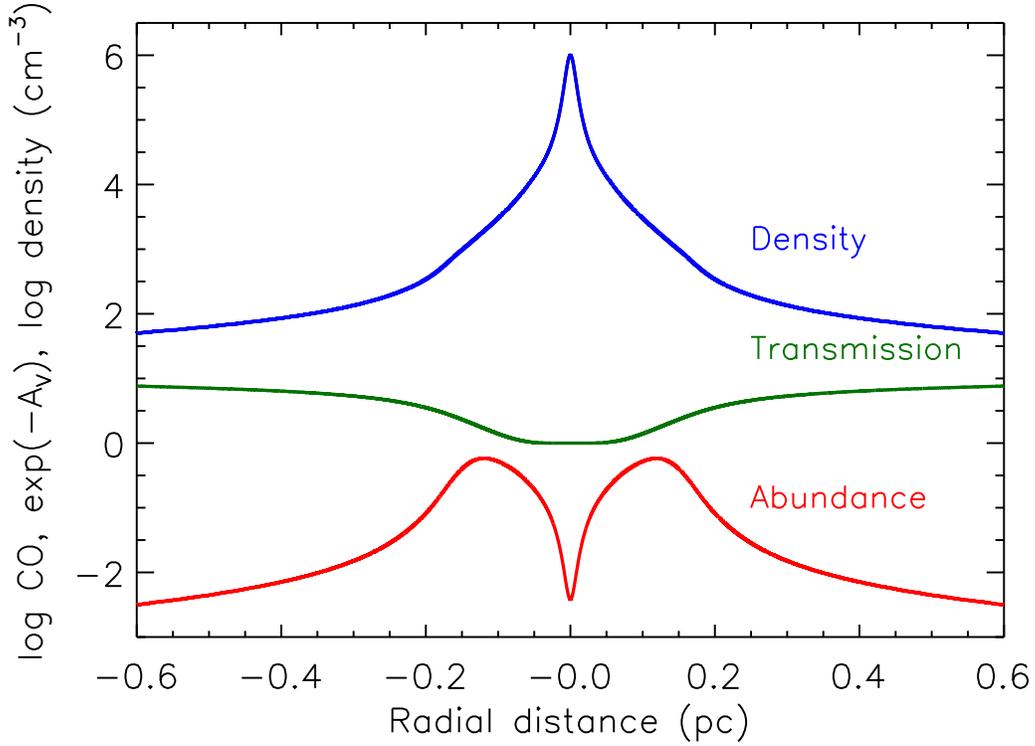} 
\caption{ Structure of a thermally super-critical core. The figure shows
the CO abundance, mean extinction, and gas density of a model cloud with
the parameters of a thermally super-critical core.
The model cloud is a Bonnor-Ebert sphere similar to that
in figure 5 of \citet{KetoField2005}.
The lowest curve is the log of the relative CO abundance. A value of zero
represents no reduction in CO abundance from photodissociation or depletion.
The middle curve is the transmission of starlight. A value of 1 represents
no extinction and zero represents total extinction. The top curve is
the log of the number density. Although the maximum radius on the plot
is 0.6 pc, the core would not be observed to have this extent
because of the low density and molecular abundance at large radii. 
Parameters listed in table \ref{parameterTable}.
} 
\label{fig:L1544_abundance}
\end{figure}

\begin{figure}[t]
\includegraphics[width=6in]{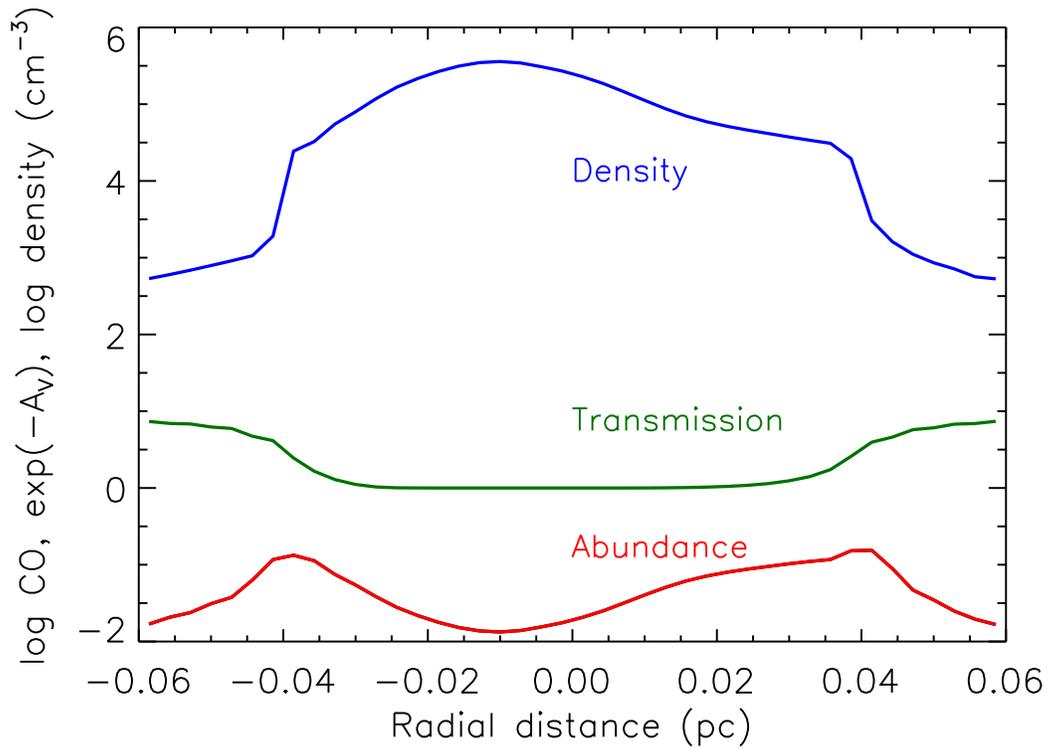} 
\caption{ Structure of a thermally sub-critical core. The figure shows
the CO abundance, mean extinction, and gas density in the same
format as figure \ref{fig:L1544_abundance}. Here the model has  
the parameters of a stable, starless, thermally sub-critical core.
The asymmetries are
perturbations from an internal oscillation. The oscillations are further
illustrated in figures 6 and 7 of  \citet{Broderick2007}.
Parameters listed in table \ref{parameterTable}.
} 
\label{fig:cloud_abundance}
\end{figure}

The combined effects of photodissociation and depletion result in the abundance of
CO peaking inside the boundary of the core where the visual extinction is just
high enough to shield the CO, but the density has not yet become high enough for
significant depletion by freeze-out. This effect is seen in figures 
\ref{fig:L1544_abundance} and \ref{fig:cloud_abundance}.
 Figure 
\ref{fig:L1544_abundance} shows a spherically-symmetric, static-equilibrium 
model with the properties
of a thermally super-critical core. This core is the same as shown in 
figure 5 of \citet{KetoField2005} but with a factor of 2 higher total mass.
Figure \ref{fig:cloud_abundance} shows a
stable, starless core 
perturbed
by an internal oscillation. 
This particular structure of an oscillating core, motivated by 
observations of B68, is the same as
illustrated in figures 6 and 7 of \citet{Broderick2007} and is included here
to show how the abundance varies in response
to an asymmetric density profile.
The sharp drop in density at the boundary of this
thermally sub-critical core 
is due to the high internal density at the boundary where the
core is truncated to give it a low mass. In B68, this 
transition implies a high pressure exterior confining gas, probably hot gas.

\section{Energy Balance and Temperature}

Models for the temperature structure of dark clouds
have been described in a number of papers
\citep{Larson1973, 
Larson1985, ClarkePringle1997, 
Evans2001, ShirleyEvansRawlings2002, Zucconi2001, 
StamatellosWhitworth2003, 
Goncalves2004, KetoField2005}.
These models include the effects of molecular
line cooling, the radiative equilibrium of dust in starlight,
dust-gas collisional coupling, and cosmic ray
heating.  We start with the model described in \citet{KetoField2005}
that includes these effects, and improve this model
by including variations in the cooling rate due to abundance 
variations and by including
photoelectric heating at the cloud boundary. 

\subsection{Line Cooling with Abundance Variations}

In a model for 
molecular line cooling  
that uses cooling coefficients
based on a local approximation such as the
large velocity gradient (LVG) radiative transfer model
\citep{Goldsmith2001}, it is straightforward to include the effect of 
abundance variations due to depletion. To do this, we use
the cooling coefficients for 
different ``depletion factors'', for example as 
listed in the tables in \citet{Goldsmith2001}. 
We assume that the abundance of coolants is given by 
our simple model of the CO abundance. 
This follows because most of the molecular line cooling is through  
transitions of  carbon species such as
$^{12}$CO, $^{13}$CO, C$^{18}$O, C, and CS \citep{Goldsmith2001}.

At the cloud edge where the CO abundance varies due to dissociation
we must also include the cooling due to C$^+$ which can be
as effective a coolant in photodissociation regions as 
CO is in molecular regions.
The C$^+$ cooling rate is given by (\citet{Tielens2005}, eqn 27),
\begin{equation}
        \Lambda_{\rm C+} = 3\times 10^{-27}  n({\rm C}^+) n({\rm H}_2)^2 
                (1 + 0.42 (X_i/1\times 10^{-3})) \exp(-92/T) {\rm \ \ ergs\ cm^{-3}\ s^{-1}}
\end{equation}
where $X_i$ is the ionization fraction, taken to be the abundance
relative to H of C$^+$, or $1.4\times10^{-4}$ C$^+$/C$_{total}$.

\begin{figure}[t]
\includegraphics[width=6in]{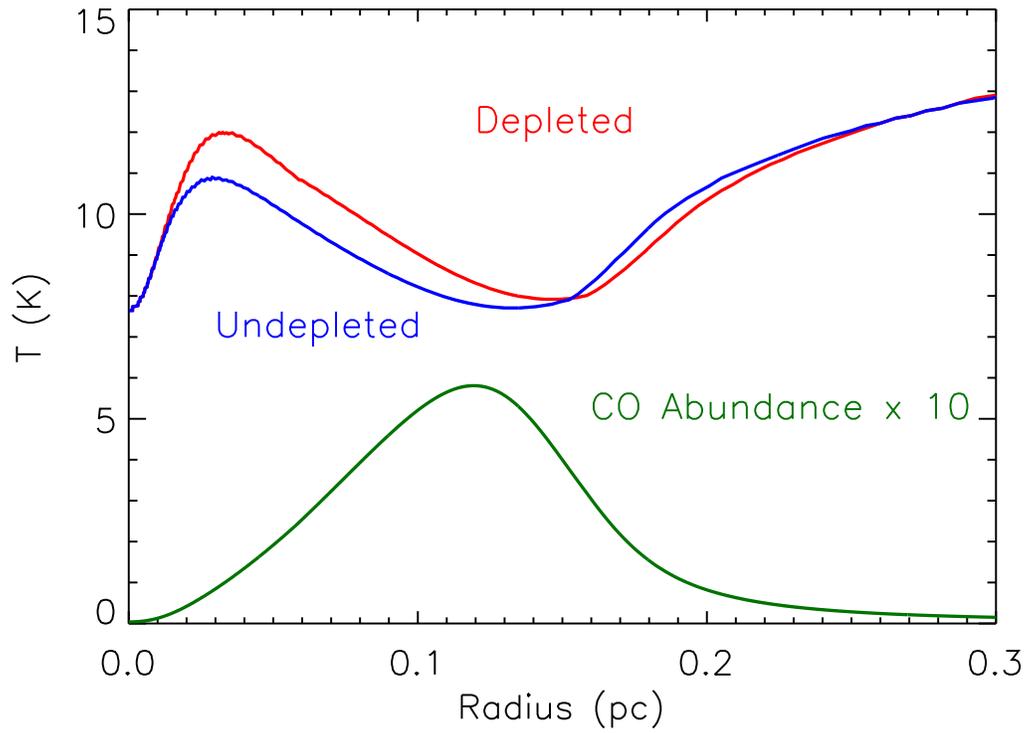} 
\caption{ 
Comparison of the temperature of thermally super-critical
cores, with 
and without variable CO abundances. Although the abundance variations
affect the molecular line cooling rate, where the abundance is most reduced
the cooling is dominated by the dust or C$^+$ rather than the molecular lines. 
The relative
abundance of CO has been multiplied by 10 so that a value of 
10 means undepleted. Parameters listed in table \ref{parameterTable}.
} 
\label{fig:temperatureL1544}
\end{figure}

\begin{figure}[t]
\includegraphics[width=6in]{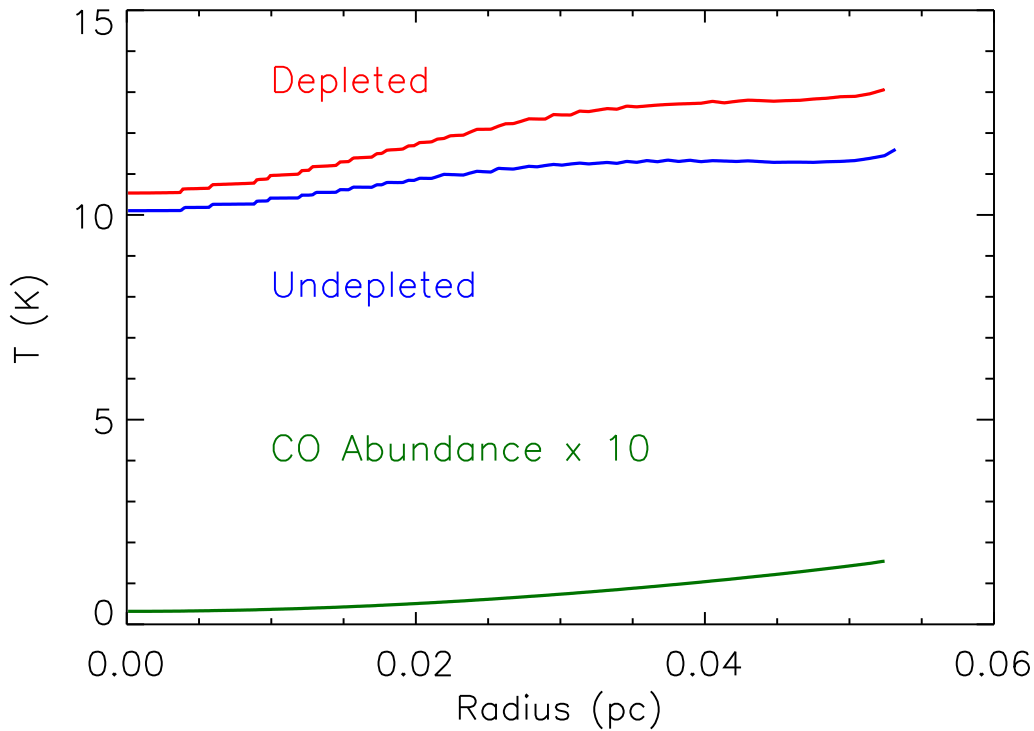} 
\caption{ 
Comparison of the temperatures of stable, starless, thermally sub-critical cores, with 
and without variable CO abundances. These cores are more nearly isothermal
than the thermally super-critical cores (figure \ref{fig:temperatureL1544}) because the
gas does not couple with the dust and is cooled primarily by line radiation.
The relative
abundance of CO has been multiplied by 10 so that a value of 
10 means undepleted. While the abundance of CO is quite low across
the core, there is still a factor of five difference between the center and
the edge. Parameters listed in \hbox{table \ref{parameterTable}}.
} 
\label{fig:temperatureB68}
\end{figure}

Figures \ref{fig:temperatureL1544} and \ref{fig:temperatureB68}
show the effect of abundance variations
on our two classes of cores. 
While  a decrease in abundance results in a decrease in the cooling rate, 
the temperatures of the depleted and undepleted cores are about the same.
In the thermally super-critical cores, 
the cooling at the core center is dominated by collisional
coupling to the dust and the cooling at the outer radii is dominated by
C$^+$. Only at mid-radii is the cooling primarily through molecular lines.
In the thermally sub-critical cores, abundance variations also do not change the
temperature too much, but for a different reason. 
The optical depths of the coolants in the thermally sub-critical cores are low enough that
the gas can always find some transitions with moderate optical depths for effective
cooling despite the changes in abundance caused by depletion. 
\citet{Pavlyuchenkov2006} also 
calculate the effect of depletion on the temperature structure of 
model starless cores.

\subsection{Photoelectric heating}

At the edge of the core, high energy photons can liberate electrons from
dust grains by the photoelectric effect. In cold dark clouds, the heating
rate is \citep{BakesTielens1994, Young2004},
\begin{equation}
\Gamma_{pe} = 10^{-24} \epsilon G_{pe}(r) n({\rm H}_2){\rm \ \ ergs\ cm^{-3}\ s^{-1}}
\end{equation}
where the efficiency factor $\epsilon = 0.5$ for the conditions in cold
dark clouds.  The factor $G_{pe}$ is the number of high energy photons ($h\nu > 6$ eV)
normalized by
the number in the general interstellar radiation field. In computing $G_{pe}(r)$, the 
intensity is averaged over frequency and direction, $\Omega$,
\begin{equation}
G_{pe}(r) = \bigg(\int^{4\pi}_0 \int^\infty_{6 {\rm eV}}J_\nu d\nu d\Omega\bigg)^{-1}
	\int^{4\pi}_0 \int^\infty_{6 {\rm eV}} J_\nu\exp(-\tau_\nu(r,\omega)) d\nu d\Omega
\end{equation}
Here $\tau_\nu(r,\omega)$ is the frequency dependent optical depth from $r$ to the
cloud surface $r=0$ along some particular direction, $\omega$.

\begin{figure}[t]
\includegraphics[width=6in]{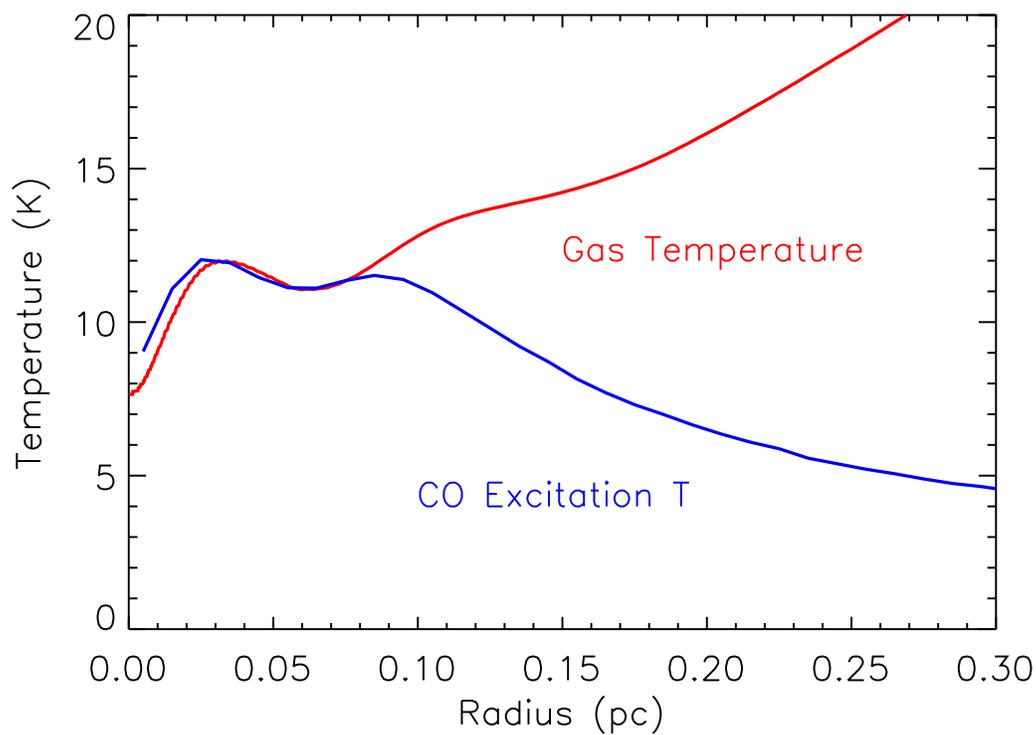} 
\caption{ 
Comparison of the gas temperature 
and the excitation temperature of CO(1-0) in an unstable, pre-stellar, thermally super-critical core.
Because the gas density (figure 3) and abundance of CO (figure 5) are so low in the outer core,
the excitation temperature of CO declines with increasing radius even as the
gas temperature increases. In the center of the core, the CO lines are optically thick and
the excitation temperature is approximately the same as the gas temperature.
Parameters listed in table \ref{parameterTable}.
} 
\label{fig:photoelectricL1544}
\end{figure}

\begin{figure}[t]
\includegraphics[width=6in]{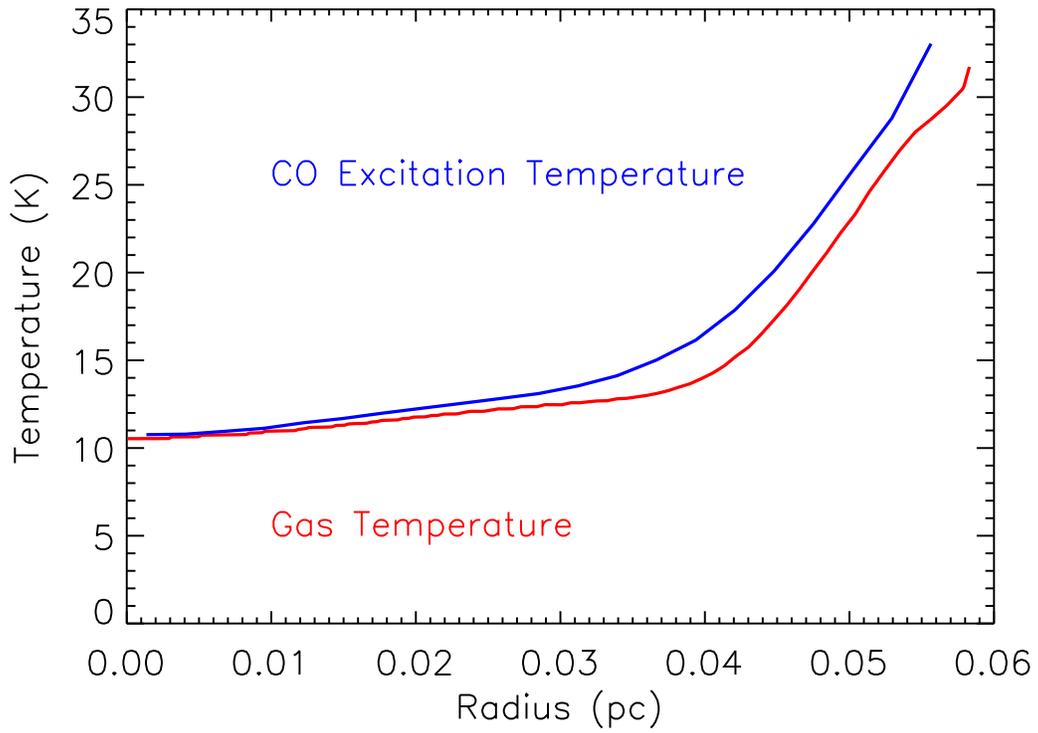} 
\caption{ 
Comparison of the gas temperature and the excitation 
temperature of CO in a stable, starless thermally sub-critical cores.
Because of the low mass of this model (1 M$_\odot$), the core is truncated at a high
enough density that the excitation temperature of the CO remains close
to the gas temperature up to the boundary. Parameters listed 
in table \ref{parameterTable}.
} 
\label{fig:photoelectricB68}
\end{figure}

\begin{figure}[t]
\includegraphics[width=6in]{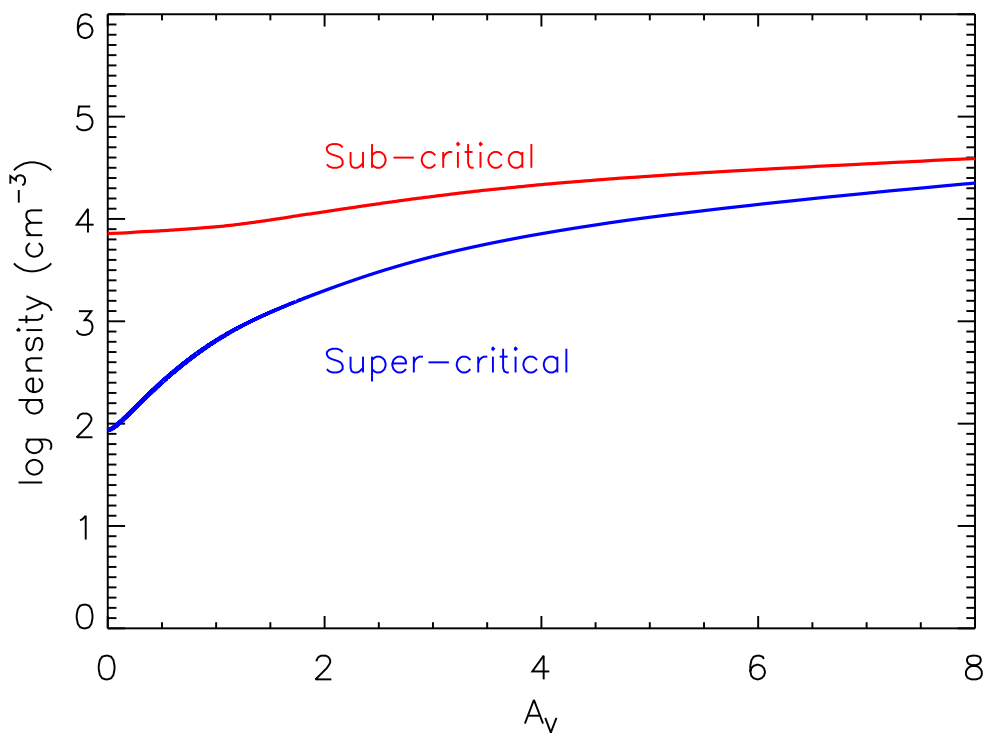} 
\caption{ 
The density structure of unstable, pre-stellar, thermally super-critical cores, 
compared to the
density of stable, starless
thermally sub-critical cores.
The sub-critical cores are truncated at a high density as would be appropriate
for low mass cores in a high pressure environment. The density of the thermally
super-critical cores decreases to a much lower value at the edge, consistent with a
lower surrounding pressure.
Parameters listed in table \ref{parameterTable}.
} 
\label{fig:twoDensities}
\end{figure}

Figures \ref{fig:photoelectricL1544} and \ref{fig:photoelectricB68} show 
the temperature structures of our two classes of cores with
photoelectric heating. Where
the gas is thin enough that shielding is ineffective, photoelectric heating 
raises the temperature.  Figures \ref{fig:photoelectricL1544} and \ref{fig:photoelectricB68}
also compare the gas temperature with the excitation temperature of CO computed
by our non-LTE radiative transfer code \citep{Keto2004, KetoField2005} assuming
a total CO abundance relative to H$_2$ of $5.625 \times 10^{-5}$
\citep{Goldsmith2001}. 
In  the thermally super-critical cores, the gas density is
very low at the boundary (figure \ref{fig:twoDensities}), and the CO is
not strongly excited by collisions.  Thus the excitation temperature of CO  declines 
with radius despite the increase in the gas temperature.
This is consistent with observations of
the diffuse interstellar medium that show H spin temperatures of about 75 K and 
CO temperatures below 5 K \citep{Burgh2007, Pineda2008}. In contrast, because of the 
high density
throughout the sub-critical cores (figure \ref{fig:twoDensities}), the CO is always
collisionally excited and the CO excitation temperature is everywhere
approximately the same as the gas temperature. The CO excitation temperature
slightly exceeds the local gas temperature at locations where the 
gas temperature is rising rapidly and the CO line has a non-neglible 
optical depth. This is because the line radiation 
from the nearby warmer gas heats the local CO level populations. 
\citet{Pavlyuchenkov2006} also calculate the brightness 
and ratios of some CO lines in models of  starless cores.

\begin{figure}[t]
\includegraphics[width=6in]{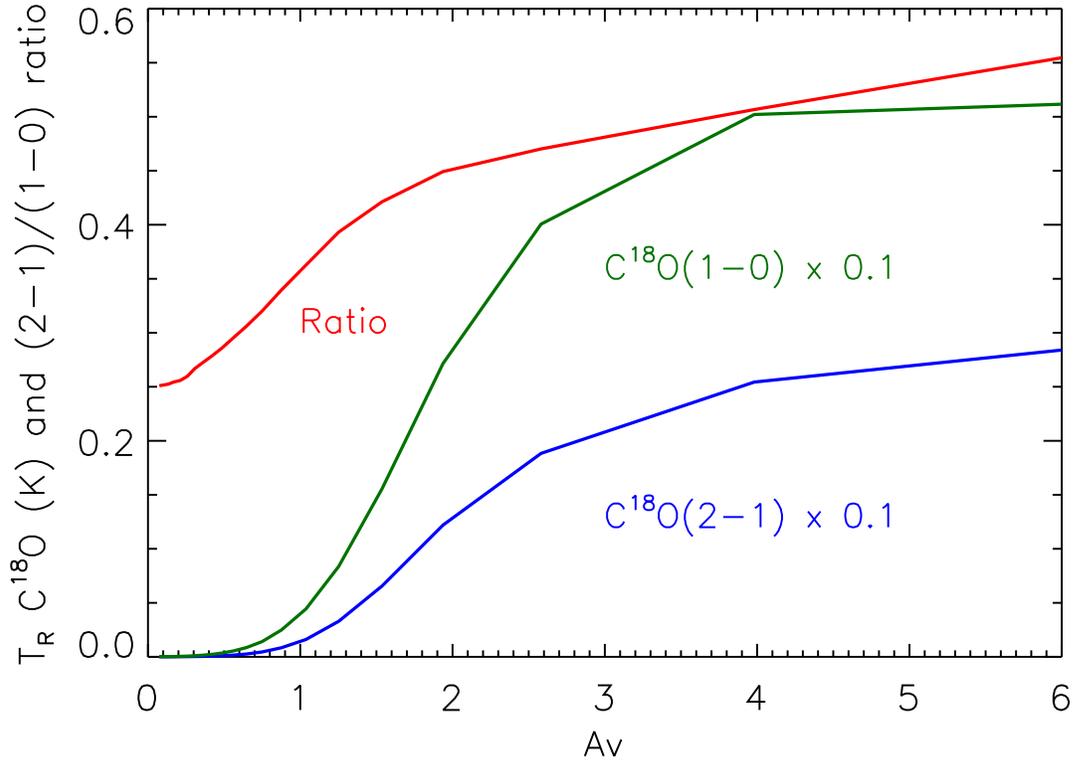} 
\caption{ 
The line intensity
of C$^{18}$O(1-0) and C$^{18}$O(2-1) (radiation temperature) and their ratio
(2-1)/(1-0) in unstable, pre-stellar
thermally super-critical cores.
The temperature structure for this model is shown in figure \ref{fig:photoelectricL1544}.
The density structure is shown in figure \ref{fig:twoDensities}.
The ratio of the CO lines is a function of  the excitation temperature of the CO. In this
model because of the low densities at the edge of the core, 
the CO excitation temperature is much lower than the gas temperature.
Parameters listed in table \ref{parameterTable}.
} 
\label{fig:c18oRatios}
\end{figure}

\begin{figure}[t]
\includegraphics[width=6in]{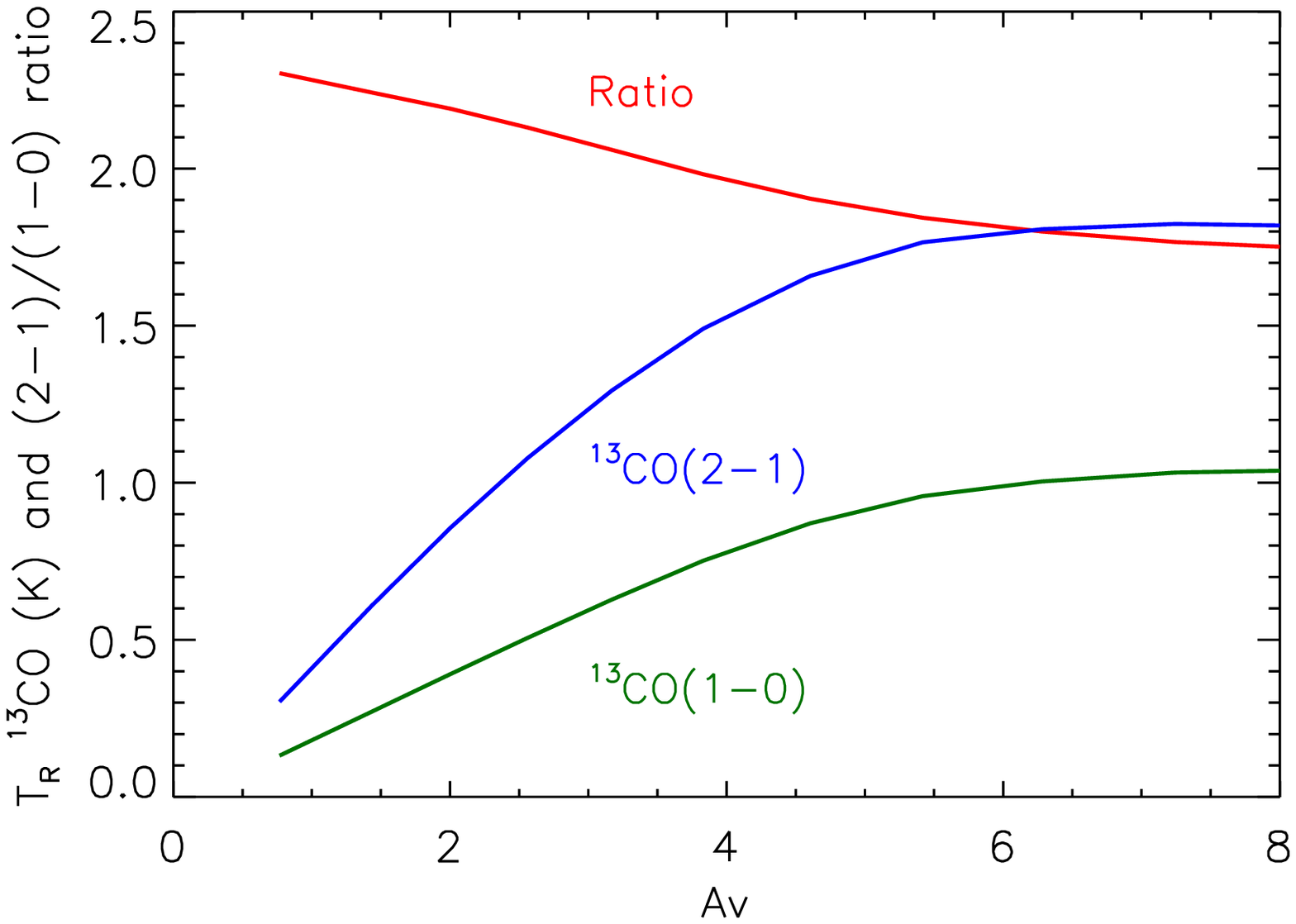} 
\caption{ 
The line intensity 
of $^{13}$CO(1-0) and $^{13}$CO(2-1) (radiation temperature) and their ratio 
(2-1)/(1-0) in
stable, starless
thermally sub-critical cores.
The temperature structure for this model is shown in figure \ref{fig:photoelectricB68}.
The density structure is shown in figure \ref{fig:twoDensities}.
The ratio of the CO lines is a function of the excitation temperature of the CO. In this
model because of the high densities at the edge of the core, 
the CO excitation temperature is always close to the gas temperature. 
Parameters listed in table \ref{parameterTable}.
} 
\label{fig:c13oRatios}
\end{figure}

Observationally, the excitation temperature of CO in cores may be determined by
comparing the brightness of the (2-1) and (1-0) transitions. Figures \ref{fig:c18oRatios}
and \ref{fig:c13oRatios} show that the ratio of CO(2-1) to CO(1-0)
declines at the edge of the thermally super-critical cores as 
the excitation temperature decreases whereas in the thermally sub-critical cores, 
the ratio rises at the edge. 
This predicted decrease in the CO excitation temperature has been observed
at the edge of TMC-1C \citep[figure 12 of ][]{Schnee2007}, a 
core whose properties put it in the  thermally super-critical class. In contrast, at the
edge of the sub-critical core B68, the ratio of the CO lines is observed to be 
constant or possibly slightly rising 
\citep[figure 5 of][]{Bergin2006}.
Although the decreasing line strengths adversely affect the signal-to-noise just at
the edge where the comparison is the most diagnostic, the observed ratios are
consistent with the different behavior of the line ratios expected in each of the
two classes of cores.   

\section{Implications}

\subsection{The rate of cosmic-ray ionization and the opacity of fluffy dust}

\begin{figure}[t]
\includegraphics[width=6in]{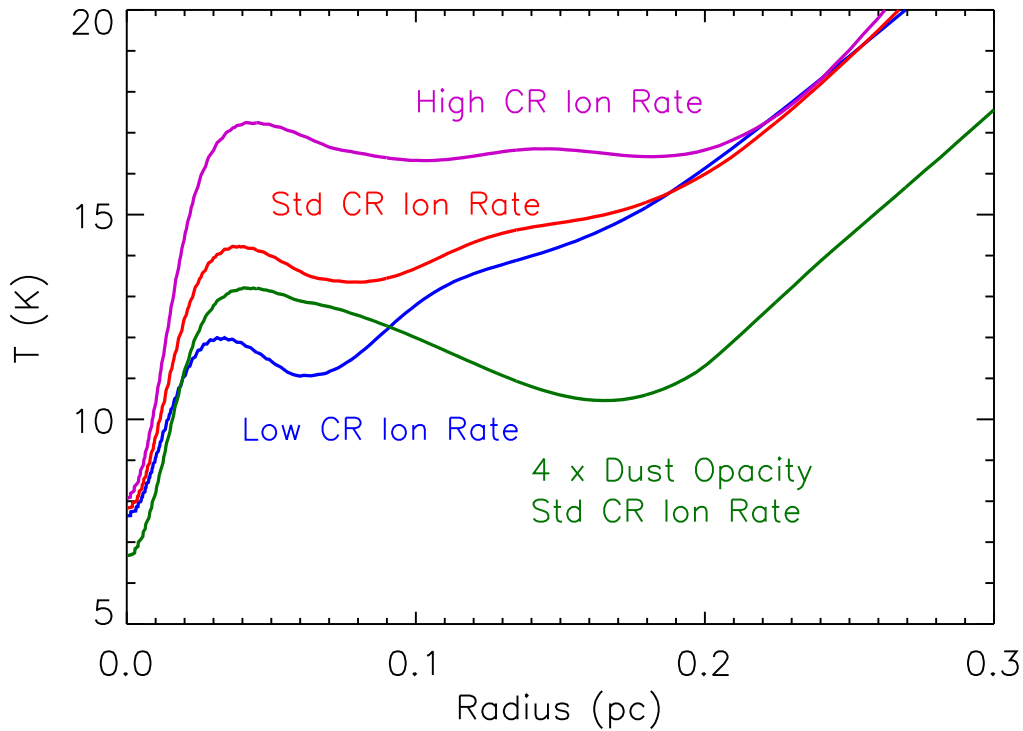} 
\caption{ 
The gas temperature in thermally super-criticals cores 
under different conditions.
The figure shows the gas temperature for a ``low'' cosmic-ray ionization rate
of $1.3\times 10^{-17}$ s$^{-1}$,  a ``standard'' rate of $3.0\times 10^{-17}$ s$^{-1}$,
and a ``high'' rate of $6.0\times 10^{-17}$ s$^{-1}$.
Because cosmic rays heat the gas directly, their rate affects the gas temperature
in the intermediate region where the cooling is not dominated by dust and the
heating is not dominated by hot photoelectric electrons. The figure also shows
the gas temperature if the dust opacity is 4 times the ``standard'' value.
This increase in dust opacity lowers the gas temperature everywhere -- in the center
and mid-radii by lowering the dust temperature, and at outer radii 
by increasing
the shielding against high energy photons that generate hot electrons.
Parameters listed in table \ref{parameterTable}.
} 
\label{fig:crRatesL1544} 
\end{figure}

\begin{figure}[t]
\includegraphics[width=6in]{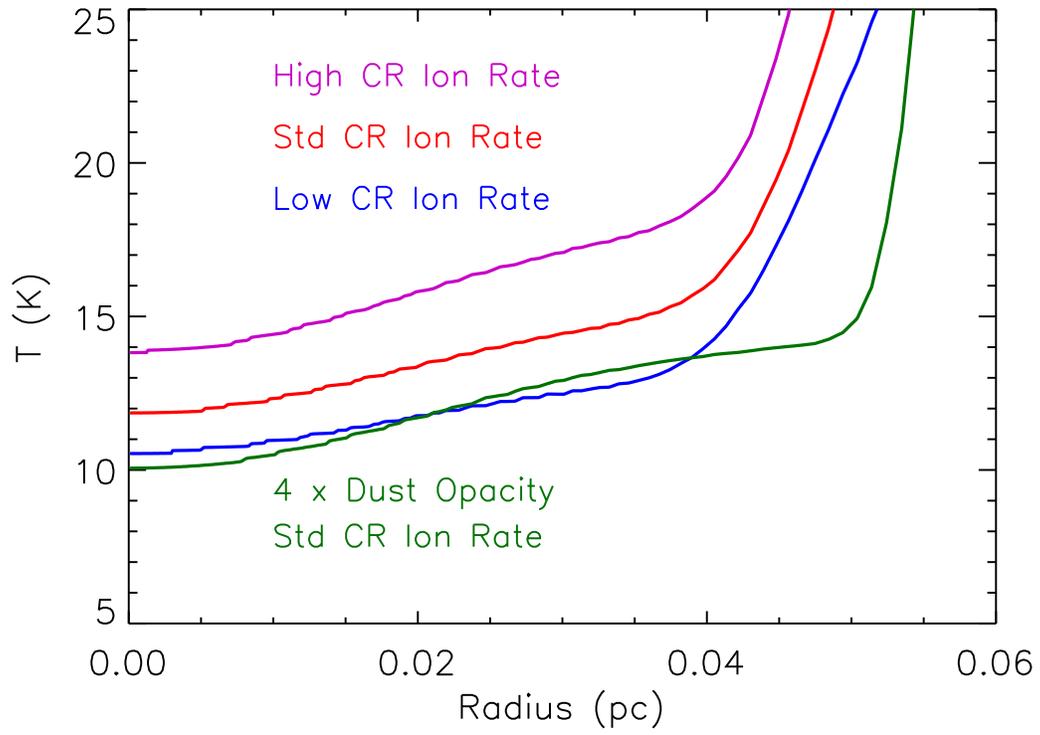} 
\caption{ 
Same as figure \ref{fig:crRatesL1544} except that the model here is
for stable starless thermally sub-critical cores, rather than super-critical.
Parameters listed in table \ref{parameterTable}.
} 
\label{fig:crRatesB68}
\end{figure}

Recent observations show promise of measuring the gas temperature with
a precision of a few degrees.
Such precision might allow a determination of
the rate of cosmic-ray ionization and also the
opacity of the dust.
In dark clouds and dense cores, observations of molecular ions as well 
as measurements of kinetic temperature limit the possible values of the 
cosmic-ray ionization rate to between 
about $1 \times$10$^{-17}$ and $6 \times$10$^{-17}$ s$^{-1}$ 
\citep{vanderTakvanDishoeck2000, Dalgarno2006}. In our models we
compare three different rates,
$1.3\times 10^{-17}$ s$^{-1}$ \citep{SpitzerTomasko1968},
$3.0\times 10^{-17}$  s$^{-1}$  and $6.0\times 10^{-17}$  s$^{-1}$
which we refer to as
as low, standard, and high cosmic-ray ionization rates.
Figure \ref{fig:crRatesL1544} and \ref{fig:crRatesB68} show
the temperature structure for
these three different rates in our two classes
of cores. In the thermally super-critical cores,
the cosmic rays most strongly affect the gas temperature at mid-radii
where the energy input from cosmic
rays is significant in the thermal balance. In the core
center where the gas temperature is collisionally 
coupled to the dust, the cosmic rays have little effect because they
carry much less energy than the dust radiation. 
At outer radii the dominant energy source is
hot photoelectrically released electrons. 
The ``high'' rate produces 
temperatures at mid-radii that are higher than generally indicated by 
observations \citep{Tafalla2004, Young2004, Crapsi2007}.
However, higher sensitivity ammonia observations at core radii of 0.05 to 0.1 pc are 
needed to make a definitive statement. 
In the thermally sub-critical cores, because the temperature
is nearly uniform except at the edge, and because the gas and dust are not
well coupled, the higher cosmic ray rates simply raise the gas temperature
across the cores.

The gas temperature is also dependent on the dust opacity. From
a comparison of CO and dust observations, \citet{KruegelSiebenmorgen1994} 
found that the opacity of dust in dark clouds is
higher than in the more diffuse interstellar medium. 
In the cold 
quiescent interiors of dark clouds, interstellar dust grains acquire 
thick icy mantles that increase their opacity.  In the high density centers
of thermally super-critical cores, the ice-coated dust may coagulate
and become fluffy, further increasing the opacity (see also 
\citet{OssenkopfHenning1994}). 
Further evidence for higher dust
opacities is provided by observations by Evans et al. (2001) that show
that a standard dust opacity results in core masses that are beyond the
gravitational stability limit if the cores were modeled as Bonnor-Ebert  
spheres. If the dust opacity were higher than the standard opacities in  
the interstellar medium, then the core masses would be less, and the 
cores more stable with lifetimes longer than free-fall times and more
consistent with the numbers of cores observed. Keto et al. (2004) 
determined the 
core masses independently
of the dust opacity from observations of N$_2$H$^+$ lines
and also
found indications that the dust mass opacity appropriate
for the more diffuse interstellar medium is too low for cloud cores.

Precise measurements of the gas temperature in starless cores could provide yet another
line of evidence for or against fluffy dust. 
Figures \ref{fig:crRatesL1544} and \ref{fig:crRatesB68} compare the gas temperatures 
in our two classes of cores calculated
with the standard dust opacities of \citet{OssenkopfHenning1994} and with the opacities 
increased by a factor of four.
The increased opacities reduce the gas temperature everywhere. In  the 
center of the cores, the gas temperature is reduced because the radiative 
input to the
dust is decreased. At the outer radii, the increased dust opacity provides shielding
against the high energy photons responsible for the photoelectric heating. Although 
there are complicating factors, the temperatures determined
from molecular line observations are generally more consistent with the lower gas 
temperatures predicted by the increased dust mass opacities of fluffy dust. This is 
particularly noticeable
in the center of the thermally super-critical 
cores where the observed temperature, about 6 K,
\citep[figure 4 of ][]{Crapsi2007} is below that obtainable with 
standard dust opacities
even with the "low" rate of cosmic ray heating. 

Another way to raise the dust optical depth without changing the dust 
opacity is to raise the gas density.
If the density of the core were higher,
the dust optical depth would also be higher, even for the same mass, and the 
central temperatures therefore
lower. Our model assumes a  central density of $1\times 10^6$ cm$^{-3}$, a little
less than estimated for L1544 by \citet{Crapsi2007}. In figure \ref{fig:crRatesL1544}
we present the results for our
lower density to maintain consistency throughout our calculations. It is simple enough
to compute temperatures for models with higher densities, and the results show that
a model with a central density a little over $10^7$ cm$^{-3}$ is required to bring
the central temperature down to 6 K if we use the standard opacities 
and cosmic-ray ionization rate.  A higher central density of course implies a
higher density throughout the core, and this would not be consistent with continuum
and molecular line observations of L1544.  

Although our models assume a dust opacity that is independent of density, 
the opacity in starless
cores may be higher at 
densities above $10^5$ cm$^{-3}$ \citep{OssenkopfHenning1994}.
In order to put stringent constraints on the cosmic-ray ionization rate, 
accurate measurements of the gas temperature as well as  the
dust continuum emission at various wavelengths will be needed to   
determine the temperature and dust opacity as a function of radius and density.

\subsection{The dynamical stability of starless cores}

The starless cores are generally successfully modeled as Bonnor-Ebert spheres
with the addition of some non-thermal energy amounting to a fraction of the
total internal energy \citep{BarrancoGoodman1998, Goodman1998, Tafalla2004, Lada2008}. 
Any disagreements with observations seem resolvable by the addition
of further physical processes to the models. As suggested in our study here, 
some of these details are significant such as
non-isothermal gas, variable molecular abundances, and perturbations in velocity and density,
but so far no observations or theoretical considerations have suggested a revision of
the basic model that the cores are pressure-supported, self-gravitating clouds.

However, the Bonnor-Ebert spheres as theoretical objects have requirements that
are not necessarily met by the starless cores. As
truncated solutions of the Lane-Emden equation, the BE spheres require an
exterior bounding pressure. 
If the bounding medium is cold enough that its own self-gravity is significant, then
the combination of the BE sphere and the confining medium cannot be dynamically stable. 
However, if the external medium is hot enough
that self-gravitational forces are negligible, then the
bounding medium itself will not collapse onto the sphere and decrease its stability.
This seems to be the case for isolated cores such as B68. 
What about cores that exist in the midst of a star forming region filled with
widespread, low-density, molecular gas? 
Is the increase in temperature of the molecular gas caused by 
photoelectric heating sufficient to stabilize the surrounding molecular gas?

\begin{figure}[t]
\includegraphics[width=6in]{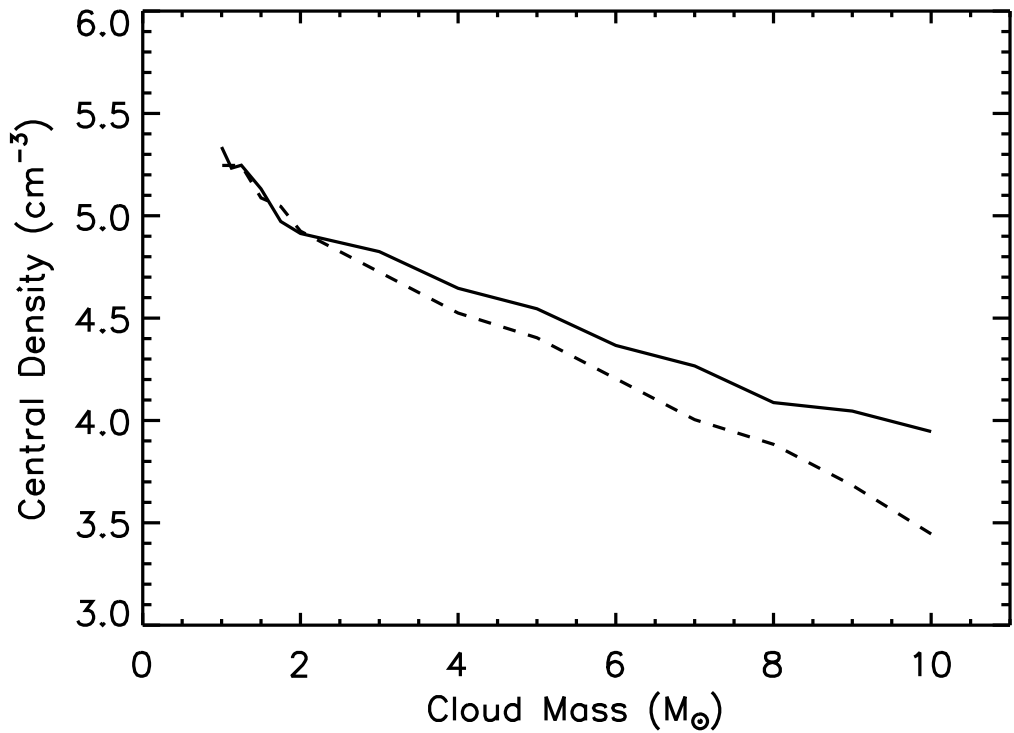} 
\caption{ 
The maximum stable central density of a starless core with (solid line) and without (broken line)
photoelectric heating at the core boundary. For the mass indicated on the abscissa, the
log of the maximum density in the center of the core is indicated on the ordinate.
The figure shows that the inclusion of the photoelectric heating at the core edge
marginally improves the stability of the more massive cores. The cores here are
modeled with the only internal energy being thermal. The addition of some non-thermal
energy in the form of microturbulence or magnetic energy would raise the curves
to higher density.
} 
\label{fig:stability}
\end{figure}

To investigate this possibility we calculate the maximum stable central densities 
for cores of different masses
with and without photoelectric heating. 
This calculation assumes the stability criterion of BE spheres \citep{Bonnor1956} applied
to non-isothermal cores. 
As a function of mass we compute the maximum or critically stable
central density which occurs at the maximum stable external pressure as indicated by
a diagram such as figure 1 of \citet{KetoField2005} or figure 1 of \citet{LombardiBertin2001}. 
Figure \ref{fig:stability} shows that photoelectric heating improves
the stability of larger cores.
However,
according to figure \ref{fig:stability} the maximum stable density
is still lower than the central densities suggested by observations.
This does not mean that observed cores are necessarily unstable.
The addition of 
non-thermal energies such as oscillations, turbulence, or magnetic fields that 
are not included in this calculation might
resolve the difference  \citep{GalliWalsmleyGoncalves2002, KetoField2005, 
Keto2006, Broderick2007, HennebelleFromang2007}. However, to include this
energy in the model we would need to know how it affects the equation of state.
If the equation of state remains as $P\sim\rho^1$, then the effect on the 
stability of the additional
non-thermal internal
energy is the same as increasing the gas temperature \citep{KetoField2005}.
The calculations for figure \ref{fig:stability} suggest that, consistent with observations, the
additional energy need only be a fraction of the thermal energy.
The analysis suggests that the increased temperature
due to photoelectric heating is
significant in improving the stability of the more massive cores, but not significant
enough to eliminate the need for some additional internal energy.

\section{Conclusions}

The structure of starless cores has been revisited, taking into account 
a simple CO chemistry and photoelectric heating in the radiative energy 
balance.  The main conclusions of this work are:

\noindent
1. Observations of the different properties of different starless cores can be
understood in the context of a simple physical model that shows different
structures and behavior depending on whether the central density is
above or below about $10^5$ cm$^{-3}$. This value depends on
other properties, in particular
masses in the range of 1-10 M$_\odot$ and
temperatures around 10 K, typical values for starless cores. 

\noindent
2. The low gas temperature ($\leq$ 7~K) recently observed toward the center
of pre--stellar cores can only be reproduced if the dust opacity is increased 
by a factor of a few, suggesting that dust grains become fluffy 
in cold dark cores where the densities are $\geq$10$^5$~cm$^{-3}$.

\noindent
3. The gas temperature of starless cores is quite sensitive to small changes
in the cosmic-ray ionization rate, $\zeta$, at volume densities below about  
10$^5$~cm$^{-3}$ (the critical density for gas--dust coupling) and at
extinction values larger than about 1~mag (where photoelectric heating 
is negligible). Thus,
high sensitivity ammonia observations of starless cores can put
stringent constraints on $\zeta$. 
Recent observations toward L1544 suggest 
$\zeta$ $\simeq$ 1$\times$10$^{-17}$~s$^{-1}$. 

\noindent
4. Photoelectric heating  affects the stability of a 
core so that it can support a higher central density or a higher 
external pressure.  However the effect is only important for the larger
starless cores which still require some additional internal energy to be
dynamically stable.

\begin{deluxetable}{cccccc}
\tabletypesize{\scriptsize}
\tablecolumns{10}
\tablewidth{0pt}
\tablecaption{Parameters of models illustrated in the figures \label{parameterTable}}
\tablehead{
\colhead{Figure}  & \colhead{core} & \colhead{Cosmic-ray}   & \colhead{Photoelectric}  
& \colhead{Depletion} & \colhead{Dust opacity}      \\
 &  & \colhead{ionization rate}   & \colhead{heating}  \\ 
} 
\startdata
3	&	YB	&	low	&	off	&	yes	&	1x \\
4	&	OM	&	isothermal & isothermal &	 	yes	&	1x\\
5	&	YB	&	low	&	off	&	both	&	1x \\
6	&	OM	&	low	&	off	&	both	&	1x \\
7	&	YB	&	low	&	on	&	yes	&	1x \\
8	&	OM	&	low	&	on	&	yes	&	1x \\
9	&	YB	&	low	&	on	& 	yes	&	1x\\
10	& 	OM	&	low	&	on	&	yes	&	1x \\
11	&	OM, YB &	low	& 	on	&	yes	&	1x \\
12	&	YB	&	low	&	on	&	yes	&	1x \\
	&		&	std	&	on	&	yes	&	4x \\
	&		&	std	&	on	&	yes	&	1x \\
	&		&	high	&	on	&	yes	&	1x \\
13	&	OM	& 	same as figure 12 \\
14	&	YB	&	low	&	on, off	&	yes	&	1x\\
\enddata
\tablecomments{YB = super-critical; OM = sub-critical; low = $1.3\times10^{-17}$ s$^{-1}$;
std = $3.0\times10^{-17}$ s$^{-1}$;high = $6.0\times10^{-17}$ s$^{-1}$; isothermal =
core structure from \citet{Broderick2007}; 1x = dust opacities of \citet{OssenkopfHenning1994};
4x = dust opacities four times higher; both = figure compares structures with depleted and
undepleted abundances.}
\end{deluxetable}

\clearpage

\end{document}